\documentclass[10pt,twocolumn,letterpaper]{article}

\usepackage{cvpr}
\usepackage{times}
\usepackage{epsfig}
\usepackage{graphicx}
\usepackage{amsmath}
\usepackage{amssymb}
\usepackage{multirow}
\usepackage{caption,subcaption}

\usepackage[breaklinks=true,bookmarks=false]{hyperref}

\cvprfinalcopy 

\def\cvprPaperID{****} 

\begin{document}


\title{Using In-Service Train Vibration for Detecting Railway Maintenance Needs}
\author{Irene Alisjahbana \\
Stanford University\\
{\tt\small alisjahbana@stanford.edu}
}
\maketitle

\begin{abstract}
    The need for the maintenance of railway track systems have been increasing. Traditional methods that are currently being used are either inaccurate, labor and time intensive, or does not enable continuous monitoring of the system. As a result, in-service train vibrations have been shown to be a cheaper alternative for monitoring of railway track systems. In this paper, a method is proposed to detect different maintenance needs of railway track systems using a single pass of train direction. The DR-Train dataset that is publicly available was used. Results show that by using a simple classifier such as the k-nearest neighbor (k-NN) algorithm, the signal energy features of the acceleration data can achieve 76\% accuracy on two types of maintenance needs, tamping and surfacing. The results show that the transverse direction is able to more accurately detect maintenance needs, and triaxial accelerometer can give further information on the maintenance needs. Furthermore, this paper demonstrates the use of multi-label classification to detect multiple types of maintenance needs simultaneously. The results show multi-label classification performs only slightly worse than the simple binary classification (72\% accuracy) and that this can be a simple method that can easily be deployed in areas that have a history of many maintenance issues. 
    
\end{abstract}

\section{Introduction}
As most of the railway infrastructure in the US is aging and nearing the end of its useful life, the need to monitor the health of the railway system is increasing. In fact, the spending on operation and maintenance have recently surpassed the spending on the infrastructure itself \cite{council}.  Today, most railway infrastructure monitoring system still rely on two methods: manual visual assessment or using a special track geometry car. However, visual inspection is known to be inaccurate in addition to being labor and time intensive. Using the track geometry car on the other hand, though very accurate, usually requires that parts of the railway be closed for maintenance.

As a result, newer and more cost efficient methods to monitor railway infrastructure health have been developed. Some examples include sensor-based monitoring systems such as optical fiber \cite{Ye2014StructuralReview} and strain gauges and image-based systems such as using SAR images \cite{Chang2017NationwideInterferometry} or employing Unmanned Aerial Vehicles (UAV) \cite{Flammini2017RailwayDrones}. Nevertheless, there are still many limitations to the aforementioned methods. For example, extremely large amount of optical fibers must be employed on the railway track to accurately monitor the whole railway infrastructure system. Image-based systems, though provides information on an interval basis, cannot provide continuous information on the condition of the railway infrastructure.

Researchers have found that the health of the railway infrastructure can be indirectly monitored through the dynamic response of a train that is in service itself. In addition to being cheaper and more reliable than visual inspection, the use of sensors that are placed on operational trains enable the continuous monitoring of the railway health. Furthermore, the availability of these sensors on the train have now provided researchers with a lot of data, thus enabling more data-driven approaches. For example, the recently released DR-Train dataset \cite{liu2019dynamic} provides the dynamic response of operational trains through vibration data and includes other corresponding data including GPS position and environmental factors. Nonetheless, past work using train vibration for railway track monitoring have mostly focused on detecting changes of the railway track and not the underlying reasons causing them. 

To that end, this paper aims to study how the different maintenance needs of railway tracks can be detected using a single pass of train vibration. In particular, this paper addresses the following questions: (1) which features and acceleration direction that is best at capturing the characteristics of the different maintenance needs and (2) how to detect multiple maintenance needs simultaneously.  

\section{Related Work}

In 2017, Lederman et al. \cite{lederman2017track} used uniaxial acceleration data obtained from sensors on board operational trains to detect 2 different changes on the railway track, particularly from track replacement and tamping, two common maintenance procedures of railway track systems. In their study, they found that the two main challenges of using train acceleration data are that the speed of the train varies from pass to pass, and that the position of the train is not known precisely as a result of the uncertainties of the GPS system. As a result, Lederman et al. aligned the raw acceleration spatially to reduce the uncertainty caused by the variability in train speed. Using this spatially aligned acceleration data, they observed that the signal-energy best represents track change which is visible after multiple passes.

In a subsequent study, Lederman et al. \cite{lederman2017trackb} proposed a novel technique to reduce the errors caused by the uncertainty of the GPS position, a major challenge that was identified in their earlier study. Here, Lederman et al. used the roughness and bumps that are found on the track profile to align the measurements. The inverse problem was solved by modelling the train as a simple damped oscillator and constraining the problem to have a limited number of bumps.

Lederman et al. \cite{lederman2017data} also presented a technique to combine features obtained through multiple sensors in order to achieve better detection accuracy. They used a two-step approach in which the data is first aligned based on the GPS position before using a novel adaptive Kalman filter that weighs the data based on its reliability to fuse the data together. By using this method, their results indicated that fusing data from multiple trains enable the change to be detected earlier. 

Liu et al. \cite{liu2019detecting} proposed an alternative method to reduce the errors caused by the uncertainty of the GPS position. Given the ground truth track coordinates obtained from the authorities, the raw data were aligned through the iterative closest point (ICP) algorithm to reduce the error of of the GPS positions. Using a variational autoencoder (VAE), they used the aligned data to detect anomalies in the signal energy of slopes of the track geometry. 

Even so, the studies done by Lederman et al. \cite{lederman2017track, lederman2017trackb, lederman2017data} and Liu et al. \cite{liu2019detecting} both focused on detecting the change itself and not the underlying reasons causing it, particularly whether the changes in the signal correspond to a type of damage that occurred or some maintenance need that needs to occur. In addition, only a single type of maintenance work or anomaly can be detected at a single time.

\section{Dataset}
\subsection{Dataset Description}
\label{dataset}
The DR-Train dataset is a public dataset first released by Liu et al. \cite{liu2019dynamic}. The dataset contains measurements obtained from 2 Light Rail Vehicles in Pittsburgh, Pennsylvania. In addition to vibration data collected by accelerometers, the dataset also includes the GPS positions, corresponding environmental conditions (temperature and weather) as well as ground truth maintenance records provided by the Port Authority of Allegheny County (PAAC). 

As explained in \cite{liu2019dynamic}, the complete data management system consists of the sensing module, data-acquisition module and data-storage module and data-processing module. 

The sensing module that was used to obtain the measurements are as follows:
\begin{itemize}
\item LRV 4306: two uni-axial accelerometers inside the cabin of the train (VibraMetrics 5102) and a tri-axial accelerometer (PCB 354C03) on the central wheel truck
\item LRV 4313: two uni-axial accelerometers (VibraMetrics 5102) and two tri-axial accelerometers (PCB 354C03)
\item Both LRV's were instrumented with a BU 353 GPS antenna
\end{itemize}{}

Detailed placement of the sensors on the two LRV's can be seen in \cite{liu2019dynamic}.
    
\begin{figure}[h!]
\begin{center}
    \includegraphics[width=3in]{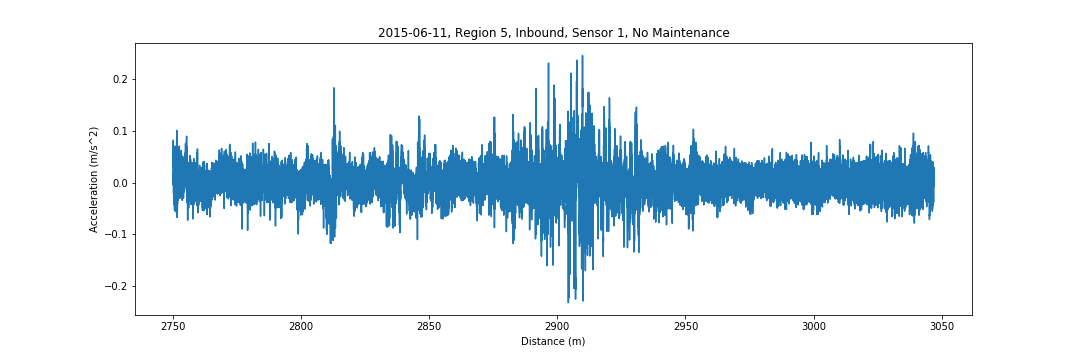}
\end{center}
    \caption{Sample raw acceleration data from sensor 1 of LRV 4313}
    \label{sample_acc}
\end{figure}

For the data-acquisition module, the National Instruments LabView was used. The system samples acceleration signals at 1.6kHz whereas the GPS position was sampled at 1 Hz. The data was downloaded manually to an external hard drive every two weeks. A sample of the raw acceleration data obtained from sensor 1 of LRV 4313 can be seen in Figure \ref{sample_acc}.

\subsection{Location of Interest}
One major issue encountered in this study is related to the ground truth maintenance logs. As explained in Section \ref{dataset}, the raw acceleration data also contains the corresponding GPS coordinates. The ground truth data however, only contains the descriptive locations and not the exact GPS coordinates (for example, Panhandle Bridge 1A). Furthermore, the location of interest had to be subjected to multiple types of maintenance work. However, the maintenance record shows that only a few locations have multiple type of maintenance work performed. 

\begin{figure}[h!]
\begin{center}
    \includegraphics[width=3in]{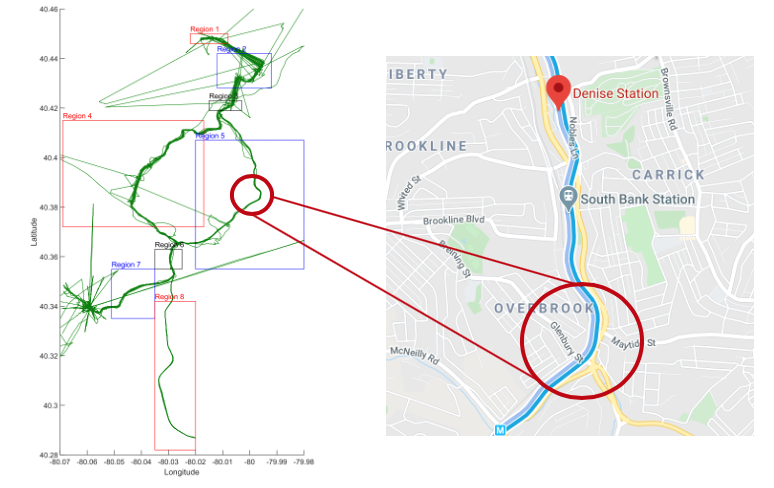}
\end{center}
    \caption{Portion of track chosen for samples as compared to Google Maps location}
    \label{location_interest}
\end{figure}

As a result, the locations provided in the maintenance record had to be manually matched with Google Maps. Taking into account the number of maintenance types performed, the location of interest selected for this project is a location with the description: Denise crossover and Glenbury crossover. This location has a recorded maintenance work of tamping and surfacing. The location of interest with the corresponding Google Maps for reference can be seen in Figure \ref{location_interest}. A 250m track portion near that area was used to extract the sample data.  

\subsection{Maintenance Record Ground Truth}

\begin{figure}[h!]
\begin{center}
    \includegraphics[width=3in]{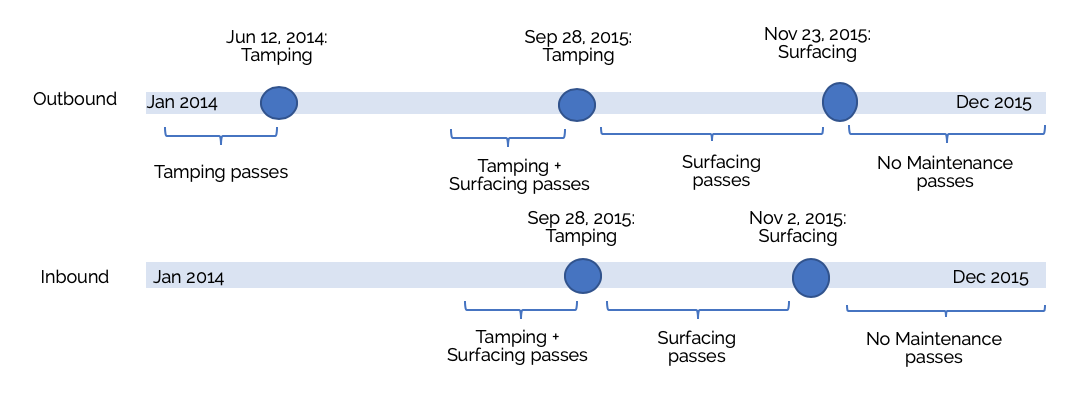}
\end{center}
    \caption{Selection of passes with respect to ground truth maintenance work}
    \label{timeline}
\end{figure}

Figure \ref{timeline} shows the dates of the specific maintenance work that was performed on both the outbound and inbound track as specified by the maintenance records.

Surfacing refers to the maintenance of the tracks in order to have the correct surface elevation, such as raising or lowering the track structure. On the other hand, tamping refers to the work that is done to pack the track ballast underneath the track itself.

Train passes that occurred within a two-month range before the maintenance work occurred were chosen for the data samples. For each sensor, a total of 17 passes were obtained for tamping, 14 passes were obtained for surfacing, 29 passes for both tamping and surfacing, and 27 passes for no maintenance work. 

\section{Methods}
\subsection{Pre-Processing}

\begin{figure}[h!]
\begin{center}
    \includegraphics[width=3in]{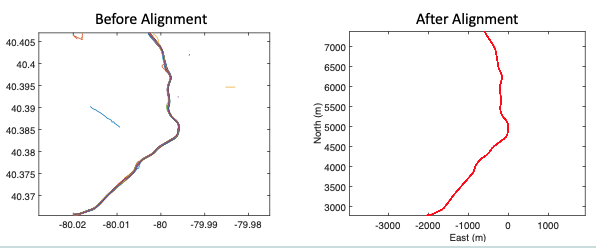}
\end{center}
    \caption{The passes before alignment and after alignment using the iterative closest point (ICP) algorithm. Passes that have bad GPS data are removed.}
    \label{icp}
\end{figure}

For the data pre-processing, the passes are first aligned with the ground truth tack coordinates using the ICP algorithm according to the method proposed by Liu et al. \cite{liu2019detecting}. Passes that have incorrect GPS positions are removed. The results of the GPS alignment can be seen in Figure \ref{icp}.

The acceleration data are then aligned spatially to reduce the uncertainties caused by the variability in train speed, similar to what was done in Lederman et al. \cite{lederman2017track}.

For all experiments, the dataset is split into training and validation sets with 80\% to 20\% ratio. 

\subsection{Feature Extraction}
\label{feature_extract}
For the individual sensor directions, the signal-energy that was first proposed in \cite{lederman2017track} is used as the features. It has been shown that this feature is more robust to position uncertainty for different types of maintenance works. The feature can be written as follows: 

\begin{center}
    $f_n = \Ddot{u}_n^2|_{x_\epsilon}$
\end{center}{}

where $\Ddot{u}_n^2$ is the raw acceleration data and ${x_\epsilon}$ indicates the acceleration of the data given the position of the train (spatially interpolated as opposed to temporally interpolated). 

Following the methods done by Lederman et al, the signal is first averaged using a 25m window of track. Then, only the 50 most discriminative indices are selected. This was done by taking the indices with the greatest difference with the mean signal.

\begin{figure}[h!]
\begin{center}
    \includegraphics[width=2.5in]{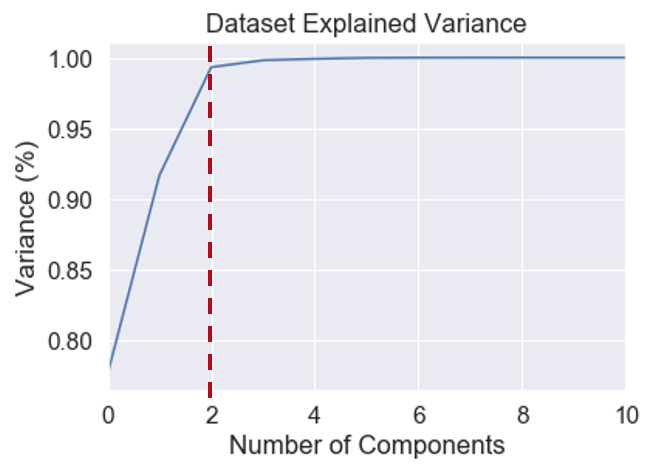}
\end{center}
    \caption{PCA results of the explained variation vs number of components for the signal energy feature}
    \label{pca}
\end{figure}

Given the limited number of data samples in the training and test sets, principal component analysis (PCA) is then performed in order to reduce the dimensionality of the features. The explained variance for the number of components included can be seen in Figure \ref{pca}. It can be seen that using only 2 components, over 95\% of the data sample variance can be explained, and therefore only the first 2 components are used as features. 

For the tri-axial accelerometer data, two types of features are defined: concatenation and major voting. In concatenation, the 50 most discriminative indices of each direction (x-, y-, and z-axis) are first concatenated in order to obtain samples with 150 features each. PCA is then performed on the concatenated features to obtain the first 2 components. In major voting, PCA is performed for each individual sensor direction (x-, y-, and z-axis). The final classification is obtained through the major voting result of each individual sensor. 

\subsection{Machine Learning Algorithms}
\subsubsection{Binary Classification}
To address the first objective of the paper in which features and acceleration direction are studied to detect maintenance needs, binary classification (no maintenance vs maintenance) is performed. Classification models such as k-Nearest Neighbours, Logistic Regression and Support Vector Machines (SVM) are implemented. 

\textbf{k-NN}: k-NN is a simple classification model in which the data are group according to the notion of distance. A data point is classified into a specific class by taking the "k" nearest neighbours and taking a majority vote based on the class of its neighboring data points. The k-NN model works well for smaller data sets and can produce decision boundaries that are non-linear and complex. It is also a non-parametric model and does not make any assumptions of the underlying sample distribution. Given the limited number of data samples, k-NN is an ideal starting point for the classification of maintenance needs. k-NN does not perform as well in higher dimensions, but given that the features of the acceleration data has been reduced to just two components, k-NN is still applicable. 

\textbf{Logistic Regression}: Logistic regression uses a logistic function to model a binary dependent variable. Unlike k-NN, the logistic regression is only able to produce linear boundary decisions. In this study, the L2 norm is used as the penalization. Furthermore, the value C, or the regularization strength, is used as the hyperparameter. 

\textbf{SVM}: The SVM is another supervised classification algorithm that uses a subset of training points in the decision function. The advantage of SVM is that it is effective even in high dimensional cases, and more robust to outliers. Similar to logistic regression, the regularization strength C is a hyperparameter. In this study, the radial basis function (RBF) kernel is used to produce more non-linear boundaries. The kernel coefficient $\gamma$ is another hyperparameter that needs to be tuned through cross-validation. 

\subsubsection{Multi-Label Classification}

\begin{figure}[h!]
\begin{center}
    \includegraphics[width=2in]{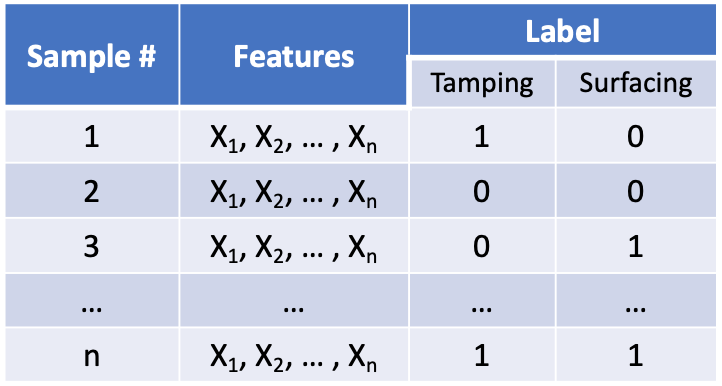}
\end{center}
    \caption{Data samples and corresponding outcome labels for multi-label classification}
    \label{multilabel}
\end{figure}

In practice, railway tracks can be subjected to multiple maintenance needs simultaneously. As a result, multi-label classification can be implemented to capture this scenario. Unlike regular classification in which each data point has one corresponding outcome label, in multi-label classification, a single data point can have multiple outcome labels (in this case tamping and surfacing). As a result, combinations of the maintenance need can be used to detect no maintenance, tamping, surfacing, or both (Figure \ref{multilabel}).

Two different methods were used to implement multi-label classification: problem transformation and adaptive algorithm. As its name suggests, in problem transformation, the classification problem is transformed from a multi-label classification into regular classification with one outcome value. This can be done through several methods, which are binary relevance, classifier chains and label powerset. In binary relevance, each label is treated as a seperate single class classification problem, and multiple models are trained for each label. In chain classifiers, the next classifier is trained on the input space and all the previous classifiers in the chain. Unlike binary relevance, chain classifier can take into account correlations between the label. Label powerset on the other hand, treats the outcomes as unique label combinations. The k-NN model is used as the basis model for all the methods. 

Adaptive algorithm on the other hand does not transfrom the problem, but instead derives regular classification methods for multi-label classification. Using the base model of k-NN, the ML-kNN \cite{zhang2007ml}, first proposed in 2007, is a classifier that is used to predict multi-label data. The ML-kNN model was implemented using the scikit-multilearn package in Python \cite{2017arXiv}.

\section{Experiments}
\subsection{Evaluation Metrics}
The evaluation metrics used to reflect the overall performance of the binary classification model is accuracy, which is the ratio of number of correct predictions to the total number of input samples. It can also be written in the following equation:

\begin{equation}
    Accuracy = \frac{TP + TN}{TP + TN + FP + FN}
\end{equation}{}

where TP = True Positives, TN = True Negatives, FP = False Positives, and FN = False Negatives. 

On the other hand, the error metric for the multi-label classification is Exact Matio Ratio (subset accuracy) and Hamming loss. The exact match ratio is the most strict metric and calculates the percentage of samples that have all their labels classified correctly. It ignores partially correct matches. The exact match ratio can be written as follows:

\begin{equation}
    Exact Match Ratio = \frac{1}{n}\sum_{i=1}^n I(Y_i = Z_i)
\end{equation}{}

where $Y_i$ is the target label and $Z_i$ is the predicted label.

The Hamming loss measures the fractions of labels that are incorrectly predicted and can be written as follows:

\begin{equation}
   Hamming Loss = \frac{1}{|N|. |L|} \sum_{i=1}^{|N|}\sum_{j=1}^{|L|} xor(y_{i,j},z_{i,j})
\end{equation}{}

where $y_{i,j}$ is the target and $z_{i,j}$ is the prediction.

The metrics are calculated for both the training set and testing set. 

\begin{figure}[h!]
\begin{center}
    \includegraphics[width=3in]{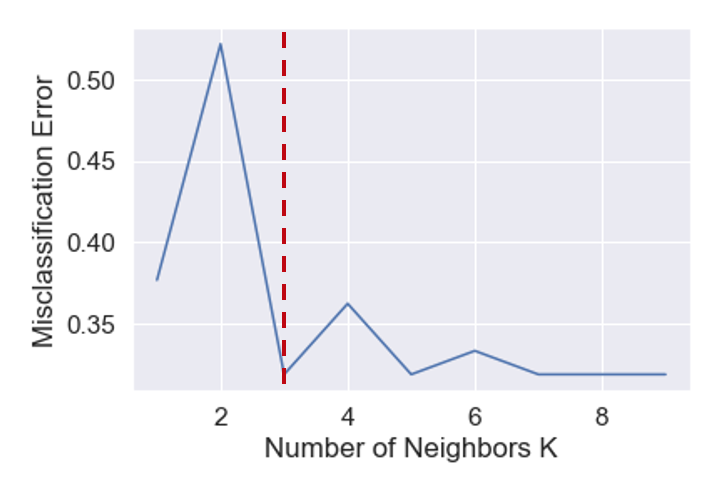}
\end{center}
    \caption{Cross validation results to determine number of k}
    \label{optimalk}
\end{figure}

\subsection{Binary Classification}

k-NN classifies a data point based on its closest neighbors, therefore, the parameter "k" will highly determine the performance of the model. Thus, the number of neighbour was tuned through using cross-validation. This is done by dividing the training set into k-folds, training it on (k-1) folds and holding 1 fold for the validation set. The accuracy is calculated for all k folds, and the mean is taken to obtain the misclassification error for the chosen hyperparameter. A sample result for the cross validation using 3-folds can be seen in Figure \ref{optimalk}. The number of folds for the cross validation was chosen by taking into account the total number of samples in the training set. 

The results for the different accelerometer directions can be seen in Table \ref{tab:direction_results}. It can be seen that the transverse direction, the direction pointing perpendicular to the railway actually performs slightly better than the vertical direction - the direction that has always been used in past studies. This however, makes sense as tamping and surfacing are maintenance works that are performed to make a smoother ride. The more unnecessary movements occurring in any other direction indicates that maintenance is needed. 

\begin{table}[h!]
\begin{center}
\begin{tabular}{|l|c|c|c|c|}
\hline
Sensors  & Optimal k & Train Acc. & Test Acc. \\
\hline\hline
z-axis (vertical)  & 8 & 0.76 & 0.73\\
x-axis (longitudinal)  & 7 & 0.84 & 0.70\\
\textbf{y-axis (transverse)} & \textbf{7} & \textbf{0.85} & \textbf{0.74}\\
\hline
\end{tabular}
\end{center}
\caption{Accelerometer direction binary classification results}
\label{tab:direction_results}
\end{table}

Table \ref{tab:acctype_results} shows the classification results for the different accelerometer types, particularly uniaxial and triaxial accelerometers. As described in section \ref{feature_extract}, two methods were implemented for the triaxial accelerometer: concatenation and major voting. It can be seen that either method performs better than the uniaxial accelerometer (only vertical direction), which further proves that the directions other than the vertical can provide more information to the maintenance needs of the railway track. 

\begin{table}[h!]
\begin{center}
\begin{tabular}{|l|c|c|c|c|}
\hline
Sensors  & Optimal k & Train Acc. & Test Acc. \\
\hline\hline
Uniaxial & 3 & 0.84 & 0.71\\
\textbf{Triaxial (Concat)} & \textbf{5} & \textbf{0.80} & \textbf{0.76}\\
\textbf{Triaxial (Voting)} & \textbf{varies} & \textbf{0.82} & \textbf{0.76}\\
\hline
\end{tabular}
\end{center}
\caption{Accelerometer type binary classification results}
\label{tab:acctype_results}
\end{table}

\begin{figure}[h!]
\begin{center}
    \includegraphics[width=3in]{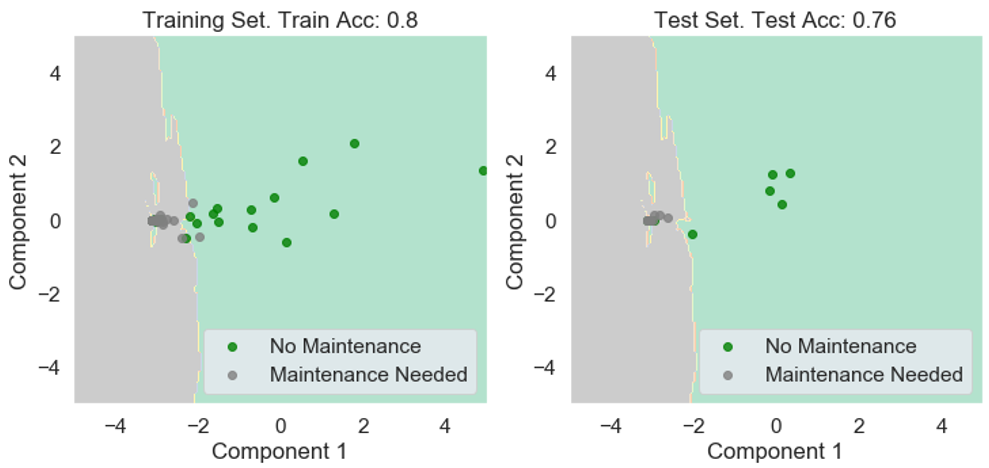}
\end{center}
    \caption{Training and test set with the decision boundaries for the triaxial accelerometer using k-NN model}
    \label{knn_decision}
\end{figure}

\begin{figure}[h!]
\begin{center}
    \includegraphics[width=2in]{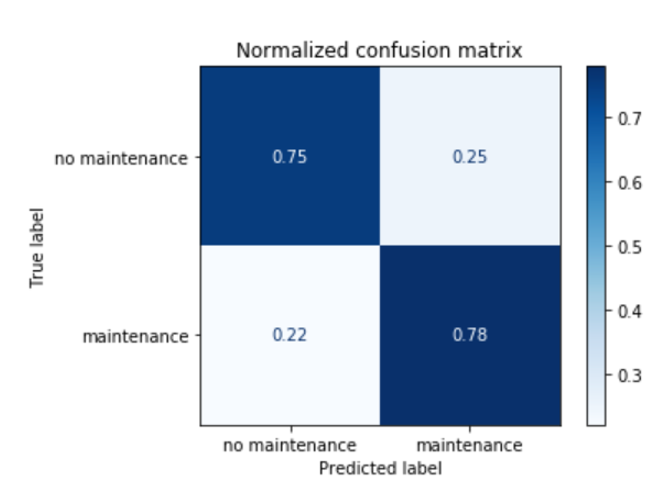}
\end{center}
    \caption{Confusion matrix for triaxial accelerometer using k-NN model}
    \label{conf_matrix}
\end{figure}

Figure \ref{knn_decision} shows the decision boundary of the k-NN model. It can be seen that the decision boundary is quite linear, with the "no maintenance" sample located on the left side of the plots. As a result, other classification models such as logistic regression and SVM with a radial basis function (RBF) kernel were implemented to improve the accuracy of the models. Similarly to the k-NN model, cross validation with 5 folds was performed to obtain the optimal parameters. 

For logistic regression model, the L2 penalty was used and the hyperparameter that had to be tuned is C, or the regularization strength. Cross validation results show that the optimal C value is 0.62. On the other hand, for the SVM model, the hyperparameters to be tuned is the C and the $\gamma$. Cross validation results show that the optimal C value is 10, with a $\gamma$ value of 0.01.

\begin{figure}[h!]
\begin{center}
    \includegraphics[width=3in]{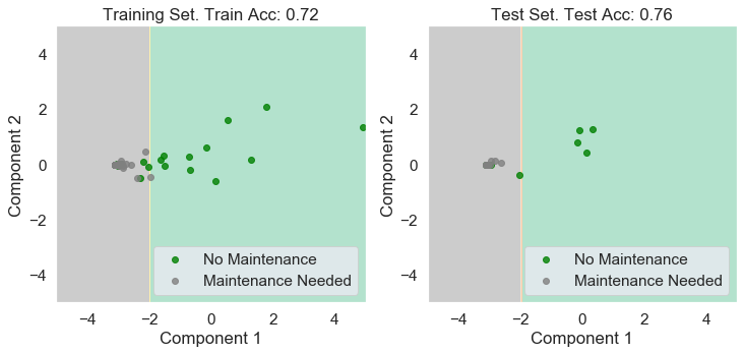}
\end{center}
    \caption{Training and test set with the decision boundaries for the triaxial accelerometer using Logistic Regression}
    \label{logreg}
\end{figure}

\begin{figure}[h!]
\begin{center}
    \includegraphics[width=3in]{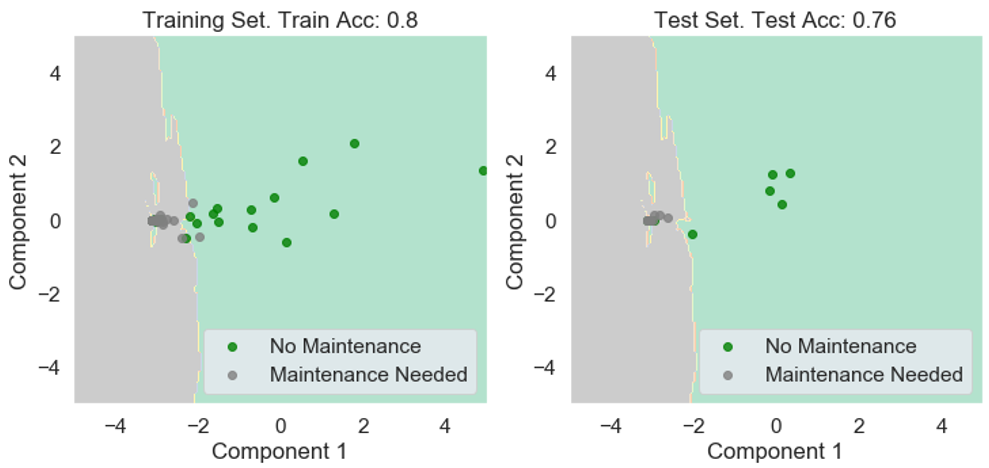}
\end{center}
    \caption{Training and test set with the decision boundaries for the triaxial accelerometer using SVM with RBF kernel}
    \label{svm}
\end{figure}

Comparison results for the three methods can be seen in Table \ref{tab:binarymodel_results}. All the method performs just as well in the test set, though the k-NN classifier had a higher train accuracy. This can be seen through the decision boundaries for the logistic regression (Figure \ref{logreg}) and SVM (Figure \ref{svm}) that are similar to that of the k-NN model (Figure \ref{knn_decision}. The higher training accuracy of the k-NN classifier can just be caused by some overfitting to the training set.

\begin{table}[h!]
\begin{center}
\begin{tabular}{|l|c|c|c|}
\hline
Model  & Train Acc. & Test Acc. \\
\hline\hline
\textbf{k-NN}  & \textbf{0.80} & \textbf{0.76}\\
Logistic Regression & 0.72 & 0.76\\
SVM RBF Kernel & 0.74 & 0.76\\
\hline
\end{tabular}
\end{center}
\caption{Binary classification results using different classification models}
\label{tab:binarymodel_results}
\end{table}

Given the uncertainty of the location of the maintenance work itself, a sensitivity analysis on the window size or length of track portion to take into account was also performed. From Table \ref{tab:lengthwindow} it can be seen that by increasing the length of track, the accuracy is reduced, and is most likely due to more noise being included. When thinking about how the method would be implemented in real life, the length of track to take into account will greatly determine the detection accuracy. 

\begin{table}[h!]
\begin{center}
\begin{tabular}{|l|c|c|c|}
\hline
Train Pass Length  & Train Acc. & Test Acc. \\
\hline\hline
\textbf{250m}  & \textbf{0.80} & \textbf{0.76}\\
500m & 0.71 & 0.72\\
1000m & 0.72 & 0.72\\
1500m & 0.76 & 0.50\\
\hline
\end{tabular}
\end{center}
\caption{Binary classification results using varying window sizes (length of track portion)}
\label{tab:lengthwindow}
\end{table}.

\subsection{Multi-Label Classification}

Results of the multi-label classification with the Exact Match Ratio error metric can be seen in Table \ref{tab:multilabel_emr_results}, whereas results using Hamming loss can be seen in Table \ref{tab:multilabel_hamming_results}. Though the Exact Matio Ratio is a very strict error metric that does not take into account partial correctness, it can be seen that the best performing model was able to achieve a 72\% accuracy, only 4\% lower than ordinary binary classification. This shows that the signal-energy feature of the train vibration is able to be used for multiple maintenance needs detection. 

Chain classifier performs slightly better overall, as it took into consideration label correlation. This would mean that a higher probability that the signal required tamping, means it also needed surfacing and vice versa. Label Powerset and ML-kNN did not perform as well probably due to the limited number of samples. 

\begin{table}[h!]
\begin{center}
\begin{tabular}{|l|c|c|c|}
\hline
Method  & Train Acc. & Test Acc. \\
\hline\hline
Binary Relevance & 0.67 & 0.72\\
\textbf{Chain Classifier}  & \textbf{0.70} & \textbf{0.72}\\
Label Powerset & 0.70 & 0.56\\
ML-kNN & 0.62 & 0.62\\
\hline
\end{tabular}
\end{center}
\caption{Multi-label classification results using triaxial accelerometer and Exact Match Ratio error metric}
\label{tab:multilabel_emr_results}
\end{table}

\begin{table}[h!]
\begin{center}
\begin{tabular}{|l|c|c|c|}
\hline
Method  & Train Loss & Test Loss \\
\hline\hline
Binary Relevance & 0.22 & 0.22\\
\textbf{Chain Classifier}  & \textbf{0.21} & \textbf{0.22}\\
Label Powerset & 0.22 & 0.31\\
ML-kNN & 0.28 & 0.31\\
\hline
\end{tabular}
\end{center}
\caption{Multi-label classification results using triaxial accelerometer and Hamming Loss error metric}
\label{tab:multilabel_hamming_results}
\end{table}

\section{Limitations and Future Improvements}

In reality, the methods that have been described in this paper will be easily applicable and trainable to new areas. However, the methods will not be able to perform real-time classification on the track signals. This is partly because of the underlying machine learning method used (kNN), which is a lazy classifier, but also because the features that was used in the model. Though the features can be easily extracted from the raw acceleration and GPS signals, the features must first be aligned based on the GPS coordinates as well as the speed of the train. This prevents the ability of real-time monitoring. 

In line with real-time monitoring, the methods described would be unable to accurately classify the signals during the transition periods. In other words, there is a grey time period between when the railway track does not need maintenance and when the track does need maintenance. This is because a specific time range (around 2 months) prior to the maintenance work was labeled as "need maintenance" and anything before that was labeled as "no maintenance". However, in reality, the threshold would be more gradual, and no clear distinction would be made. A deeper study on the effects of the time window on the accuracy of the models can be performed. Perhaps one would use passes from multiple days instead of a single day to see whether the model would consistently predict "need maintenance". Thus, the confidence levels of the classification would increase if the track portion is consistently classified as "need maintenance". 

Furthermore, only one location, the Denise Crossover - Glenbury Crossover, was used as a case study in this study. Therefore, the ability for the model to work well on other locations is still unknown. The model will most likely have to be retrained to be implemented in other locations. Finally, only two types of maintenance work, tamping and surfacing was taken into account in this study. In reality, there are many other types of maintenance work that a railway track can be subjected to. Including more locations and types of maintenance needs can be an extension to this study.  

As described in Section \ref{location_interest}, one of the main challenges of this study is the unavailability of GPS coordinates corresponding to the maintenance work. Obtaining the corresponding GPS coordinates associated with each location of the maintenance work would be extremely beneficial to generalizing the models. This might mean some additional agreement with the PAAC authorities as exact track damage locations might not be available to the public. However, if this information were available, more location and maintenance need types can be included into the training and testing dataset.

\section{Conclusion}

In this study, we demonstrated how detecting maintenance needs, particularly tamping and surfacing, of railway tracks can be performed using a single pass of in-service train vibration. 
The results show that the transverse direction was able to provide more information on the maintenance needs of railway tracks as compared to the vertical direction, the direction that has mostly been used in past studies. This was further supported by the highest accuracy obtained by the triaxial accelerometer with an accuracy of 76\% using a simple k-NN classifier. The results also show that increasing the length of track taken into consideration will drastically reduce the classification accuracy. 

Finally, this study showed that multiple maintenance needs can be detected simultaneously through multi-label classification while only reducing the overall accuracy by 4\%. 

Though this study only demonstrated the method on a single location and only two maintenance needs, it is a step towards an early warning detection system for detecting maintenance needs of railway track systems. The methods presented in this study is easily trainable and applicable to any location that have a history of many maintenance issues.

\section{Acknowledgments}
I would like to acknowledge the Stanford University teaching team of CEE 286 (Structural Health Monitoring) particularly Professor Haeyoung Noh and Mostafa Mirshekari for the constant feedback on the project.

{\small
\bibliographystyle{ieee}
\bibliography{egbib}
}

\end{document}